\documentclass[12pt]{iopart}
\usepackage{epsfig}

\begin{document}

\title{Molecular sensor based on graphene nanoribbons}

\author{L. Rosales$^{1}$, C. D. Nunez$^{1}$, M. Pacheco$^{1}$,  A. Latg\'{e}$^{2}$ and P. A. Orellana$^{1}$}

\address{$^1$ Departamento de F\'{i}sica, Universidad T\'{e}cnica Federico Santa Mar\'{i}a,
Casilla 110V, Valpara\'{i}so, Chile} 

\address{$^2$ Instituto de F\'{i}sica, Universidade Federal Fluminense, 24210-340, Nitero\'{i}-RJ, Brazil}

\ead{luis.rosalesa@usm.cl}

\date{\today}

\begin{abstract}
In this work we study thermoelectric properties of  graphene nanoribbons with side-attached organic molecules. By adopting a single-band tight binding Hamiltonian and  the Green's function formalism, we  calculated  the  transmission and  Seebeck coefficients for different hybrid systems.  The corresponding thermopower profiles exhibit a series of sharp peaks  at the eigenenergies of the isolated molecule. We study the effects of the temperature on the thermoelectric response, and  we consider random configurations of molecule distributions, in different disorder regimes.  The main characteristics of the thermopower  are not destroyed under temperature and disorder, indicating the robustness of the system as a proposed molecular thermo-sensor device.

\end{abstract}

\maketitle

\section{Introduction} \label{Intro}

Recent reports \cite{Balandin,ShuHui,Dragoman,Wang}, predict interesting changes of the electronic and thermoelectric properties of graphene-based systems, as a function of its dimensionality. The possibility of modulating and enhancing their physical responses as a function of gate potentials, disorder, defects and other types of electronic confinement makes these systems good candidates for new technological devices\cite{jarillo,pedersen}. 
Graphene nanoribbons (GNRs) are quasi one-dimensional systems which can be obtained from a graphene layer by different experimental techniques\cite{chinos,Ci,Kosynkin,Terrone}. The additional  confinement due to  presence of  edges, induces strong changes on the electronic structure and, on the electric and thermoelectric properties of the nanoribbons\cite{Cuniberti,Guo,Kosina,Jinghua}. In that sense GNRs are promised systems for functionalization and formation of new hybrid structures.

One possible application is concerned with the capability of graphene-based materials to detect molecules attached to the systems, such as nitrogen dioxide and 
trioxide, water, and different aromatic molecules. Nitrogen-based molecules act like electron acceptors or donors, depending of their size and internal structure, changing the local carrier concentration of the graphene. Step-like modifications in the resistance of the system \cite{schedin} are then detectable at room temperature,  even at a very low concentration of molecules.  On the other hand, aromatic molecules are easily detected by a graphene-base device \cite{Dong} due to the strong binding between graphene $\pi$-bonds and molecular aromatic rings. Actually, graphene sensibility is better than any material currently used in gas-sensor devices \cite{Wehling}. 
 
In previous works we have addressed the effects, on  GNRs conductance,  of organic molecules adsorbed at the ribbon edge\cite{Rosales,Rosales2}. All the considered molecule distributions were  ordered configurations.  We found that the corresponding molecule energy spectrum  is obtained as a series of Fano antiresonances in the conductance of the system, 
and we proposed that GNRs could be used as spectrograph-sensor devices.  Complementary information to conductivity  in elucidating the mechanisms dominating electronic transport properties may be provided  by  thermopower measurements.  Actually, thermopower effects have been measured in carbon-based systems\cite{Zuev}. In this work we calculate the thermopower of armchair graphene nanoribbons (AGNRs) in the presence of linear polyaromatic molecules (LPHC) attached to the ribbon edges. We calculate the Seebeck coefficient and the electronic transmission of the systems, for different molecular configurations, taking into account one molecule, a finite number of equidistant molecules and also random distributions of molecules, which certainly is a best choice for the proposed experimental scenario. We have found that the thermopower response is enhanced by the presence of the molecules.  Our results show that thermopower reflects the molecular spectra for all considered temperatures, even in the case of  random molecular configuration. This evidence suggests possible novel applications for molecules detection based on  thermoelectric  properties of  graphene nanoribbons.

\begin{figure}[ht]
\centering
\includegraphics[width=0.44\textwidth,angle=0,clip]{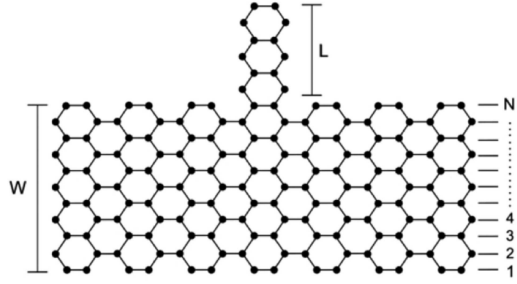}
\caption{Schematic view of the considered hybrid system composed of an armchair nanoribbon (order N and width W) and an organic molecule with L hexagons pinned at the edges of the GNR.} 
\end{figure}
\section{Theory} \label{teorico}

The thermopower or Seebeck coefficient  $S$ is defined  as the voltage drop induced by a temperature gradient at vanishing current, $S=-\Delta V/\Delta T|_{I=0}$, in the limit of  $\Delta T\rightarrow 0$.
The electric current is obtained within a single-particle picture using the Landauer approach\cite{Butcher} 
\begin{equation}
I = \frac{e}{\pi \hbar} \int_{-\infty} ^\infty   \tau(E)(f_L(E)-f_R(E))dE\,\,\,,
\end{equation}
where  $\tau(E)$ is the transmission coefficient and $f_{R,L}$ are the Fermi distributions of the right and left leads.  The thermopower is calculated in the linear response regime (i.e., $|\Delta T|<<T$ and $|e\Delta V|<<\mu$, with $\mu$ being the equilibrium chemical potential, and T the temperature), and is given by
 \begin{equation}
S(\mu,T)=\frac{1}{eT}\frac{\int_{-\infty} ^\infty \left(-\frac{\partial f}{\partial E}\right) (E-\mu) \tau(E) \, dE}{ \int_{-\infty}^\infty \left(-\frac{\partial f}{\partial E}\right)  \tau(E) \, dE}\,\,\,.
\label{Kn}
\end{equation}

\begin{figure}[ht]
\centering
\includegraphics[width=0.47\textwidth,angle=0,clip]{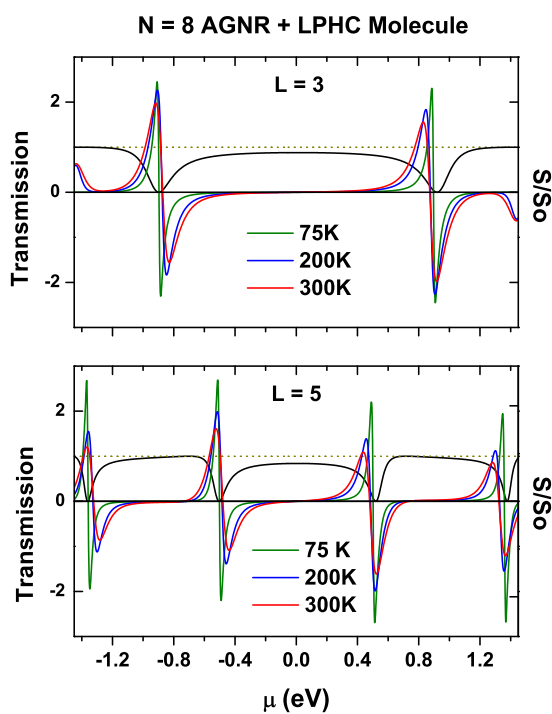}
\caption{ (Color online) Temperature dependence of the transmission and Seebeck coefficients as a function of the chemical potential for an N = 8 AGNR with a single attached  molecule. Two molecular lengths have been considered, $ L=3$ and $5$.} \label{temperature}
\end{figure}

\begin{figure}[ht]
\centering
\includegraphics[width=0.45\textwidth,angle=0,clip]{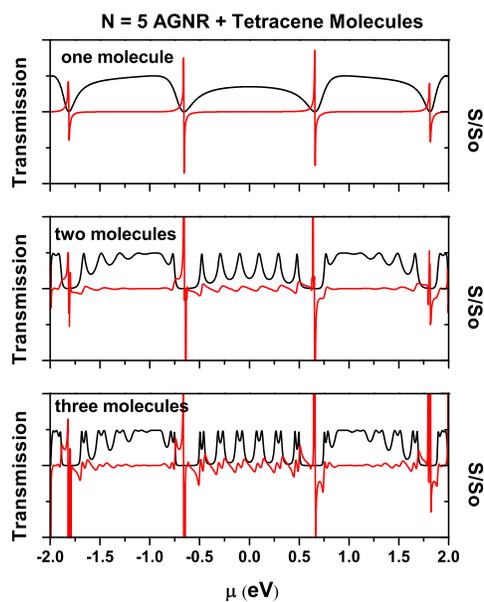}
\caption{(Color online) Transmission and Seebeck coefficients as a function of the chemical potential for a N = 5 AGNR for one, two and three tetracene molecules separated by $20$ unit cells.  } 
\end{figure}

Using the Sommerfeld expansion, the thermopower may be written as $S=(1/eT) K_1/K_0$, with
\begin{eqnarray}
K_0\!&=&\!\tau+\frac{\pi^2}{6}\tau^{(2)}\xi^2+\frac{7\pi^4}{360}\tau^{(4)}\xi^4+ O(\xi^6), \label{K0} \\
K_1\!&=& \!\frac{\pi^2}{3}\tau^{(1)}\xi^2+ \frac{7\pi^4}{90}\tau^{(3)}\xi^4+O(\xi^6),\label{K1}  
\end{eqnarray} \label{kas}
\noindent where $\xi=k_B T$ and $\tau^{(n)}\equiv\tau^{(n)}(\mu)=(d^n\tau/dE^n)(\mu)$.  In the systems we are interested, the transmission coefficients present a set of Fano antiresonances\cite{Pedro} and we have to go beyond the Mott approximation (linear in T) taking into account terms of higher order in $\xi$.

\begin{figure}
\centering
\includegraphics[width=0.48\textwidth,angle=0,clip]{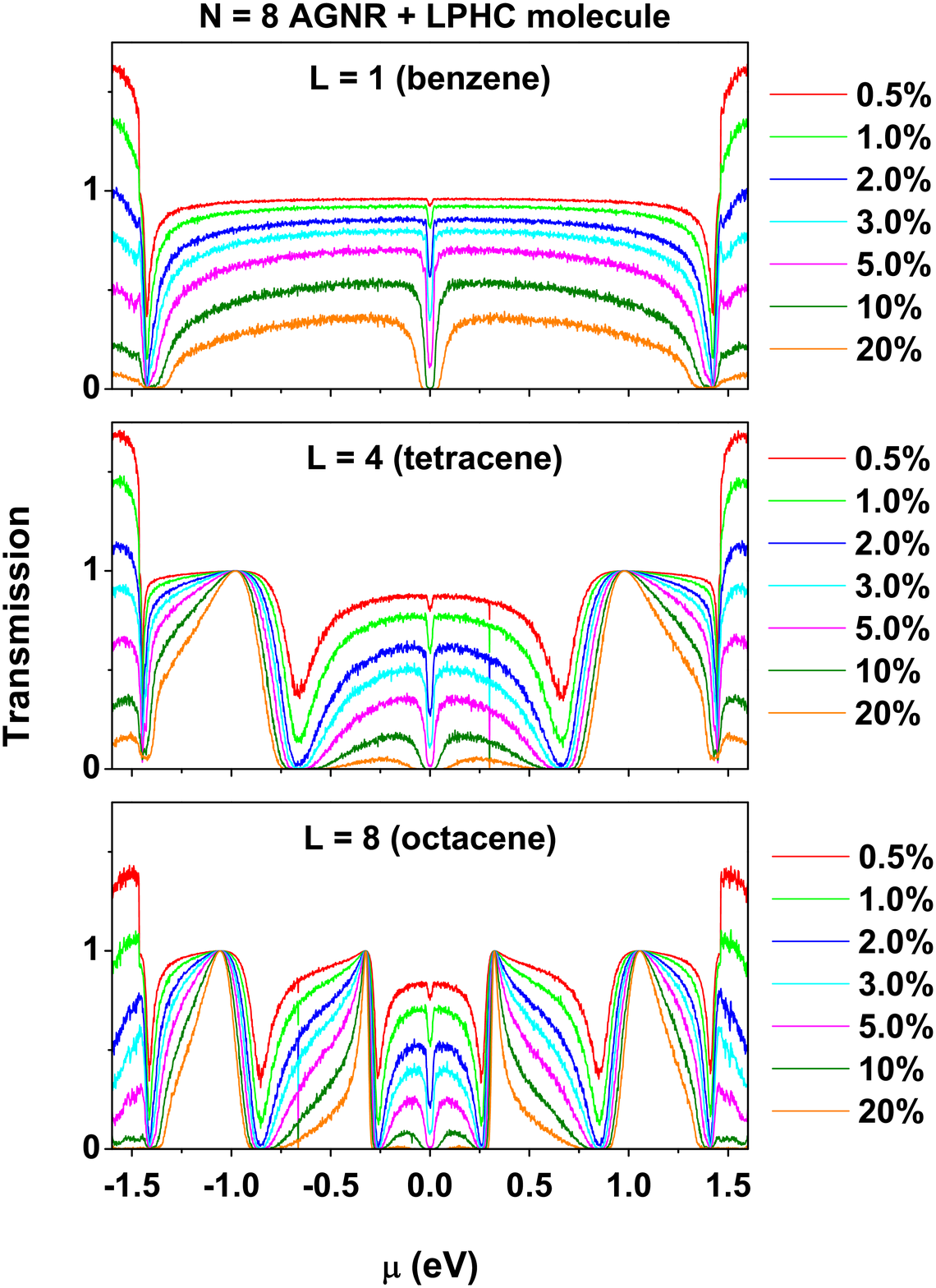}
\caption{(Color online) Transmission coefficients as a function of the chemical potential for an  N= 8 AGNR in the presence of a random distribution of  molecules with different lengths (L= 1, 4, and 8). The distinct curves represent different disorder parameters,  marked on the right of each panel.} 
\end{figure}

A single $\pi$-band tight binding Hamiltonian, taking into account only
nearest neighbour interactions with a hopping $\gamma_0 =
2.75\,eV$, is used to describe different configurations of  armchair graphene nanoribbons with edge-attached molecules. The transmission coefficients are calculated 
within the Green's function formalism given by 
\begin{equation}\label{LandauerG}
\tau\left( {E } \right) =
{\mathop{\rm Tr}\nolimits} \left[ {\Gamma _L G_C^R \Gamma _R G_C^A }\,\,
\right],
\end{equation}
where the retarded (advanced) conductor
Green's function, $G_C^{R(A)}$ are obtained following  a real-space renormalization  scheme\cite{Nardelli,Rosales} and $\Gamma_{L,R} $ describe the particle scattering between the leads and the  conductor. 

\section{Results and discussion} \label{Resultados}

\begin{figure*}[ht]
\centering
\includegraphics[width=0.46\textwidth,angle=0,clip]{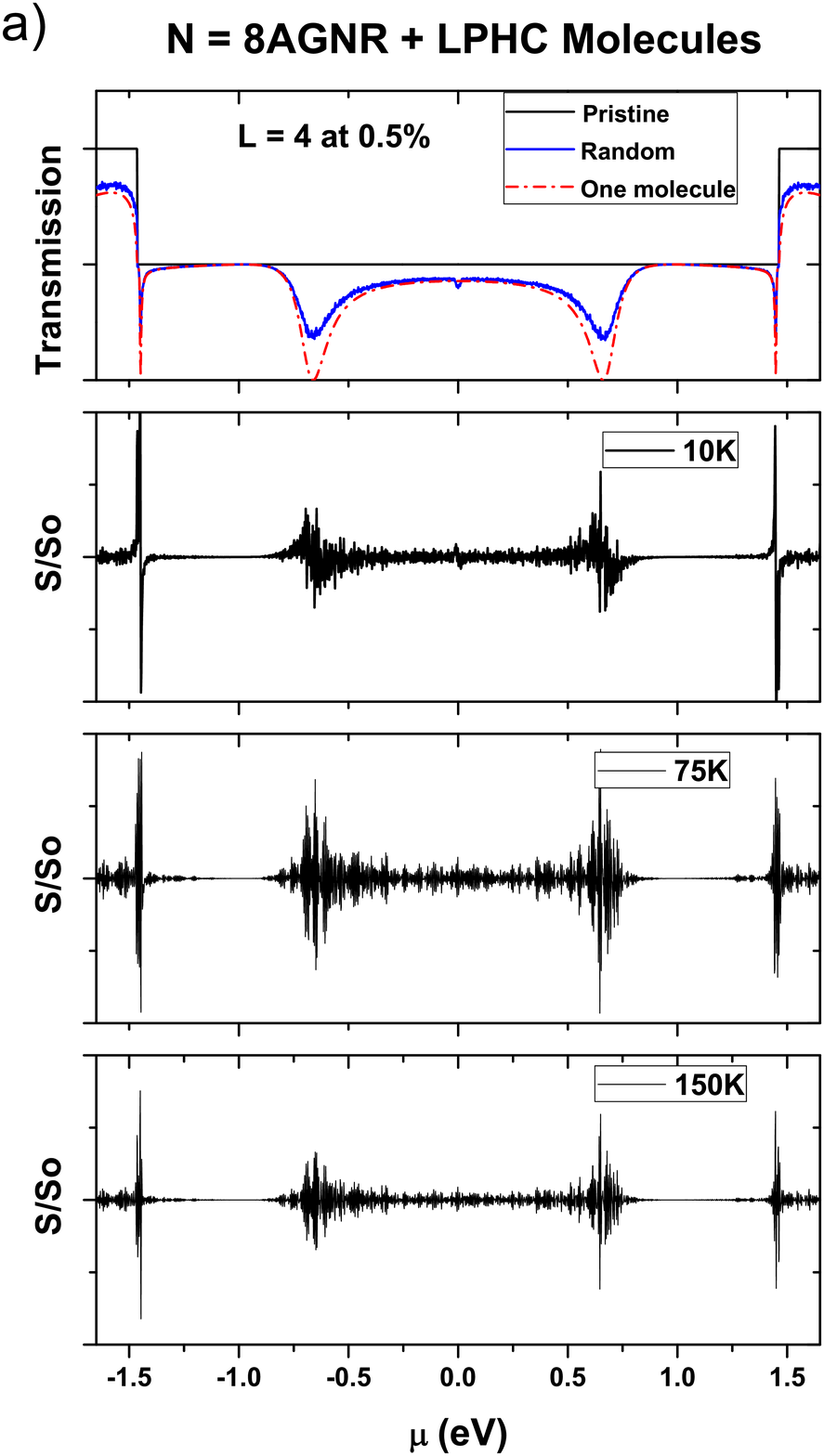}
\includegraphics[width=0.46\textwidth,angle=0,clip]{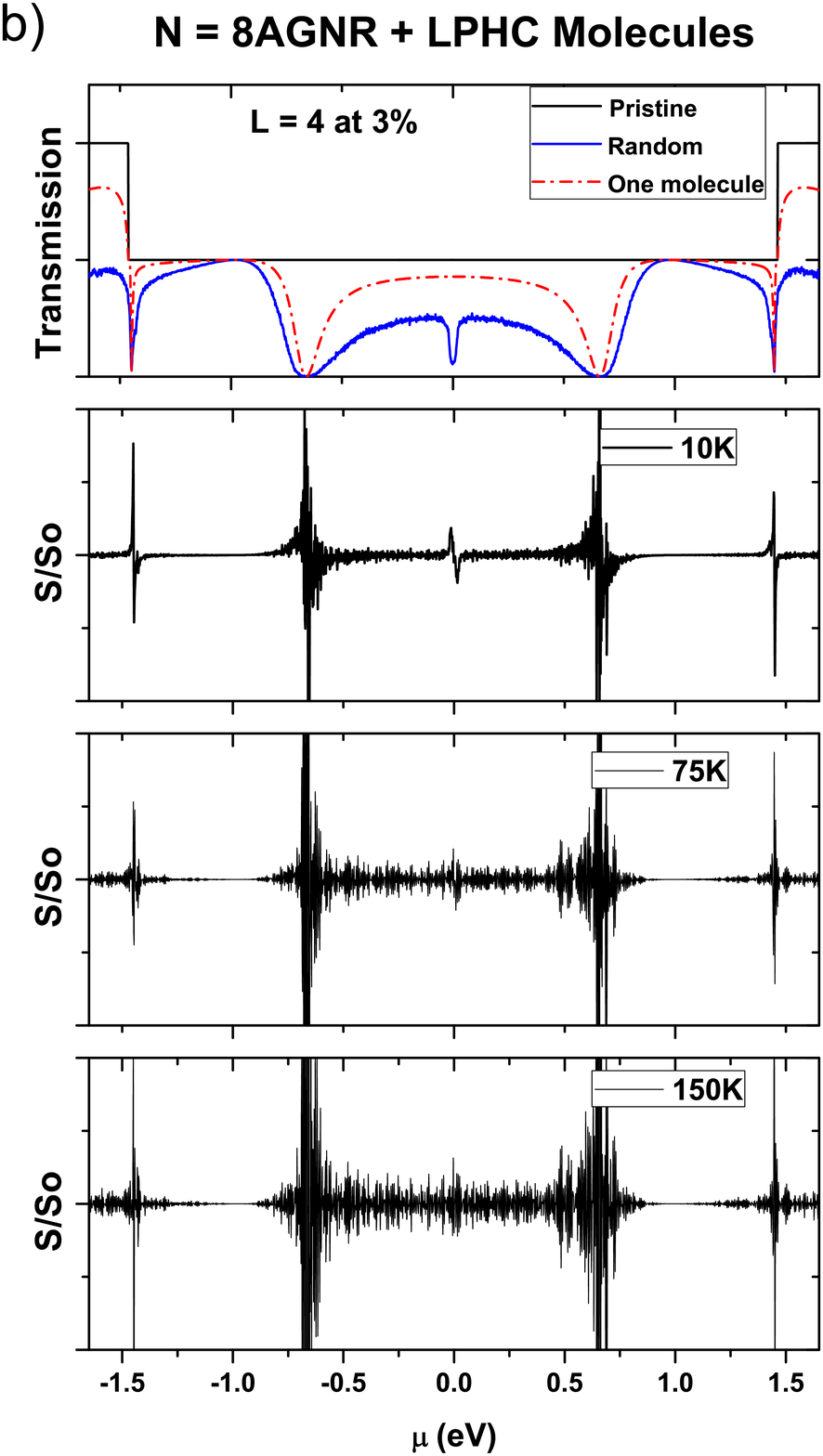}
\caption{(Color online) Transmission  and Seebeck coefficients as a function of the chemical potential for an  N= 8 AGNR in the presence of a random distribution of  molecules. Panels (a) and (b) represent the transmission and temperature dependence of the thermopower,  for a low molecular concentration and a higher concentration, respectively. } 
\end{figure*}

In what follows we present results for an N=8 AGNR with attached LPHC molecules. Figure \ref{temperature} displays the temperature dependence of the transmission and thermopower coefficients, in terms of $S_0=\pi^{2} k_B/3e$, as a function of the chemical potential. Two molecule lengths, $ L=3$ and $5$, are considered, with $L$ denoting the number of hexagons of the quasi one-dimensional molecules. Sharp structures are clearly shown in the Seebeck coefficient at the chemical potential values where Fano antiresonances in the transmission occur. As it was found previously\cite{Rosales} these antiresonances are a signature  of the  attached molecule spectrum. The Seebeck coefficients as a function of the chemical potential show similar features with results obtained for graphene\cite{Guo}.  The sharpness of the Seebeck peaks is not strongly affected by increasing temperature exhibiting just smoothed structures. As expected, the  molecule length determines the number of  signals in the termopower as a function of the chemical potential. 

In order to analyse the effect on the thermoelectric transport of the number of attached molecules to the AGNR system, we show in Fig. 3 the results of transmission and Seebeck coefficients for three different configurations. We consider the cases of one, and two and three Tetracene (L=4) molecules separated by twenty unit cells. The main features of the thermopower curves are robust against the increasing number of attached molecules. Actually, the intensity of those peaks exhibits a pronounced enhancement. These results indicate that nanoribbon systems can be proposed to design molecular sensor devices.  

 An important point to mention is the fact that the molecular distribution along the ribbon edges does not destroy the main signatures of the transmission determined by the type of molecule attached. We have considered an N = 8 AGNR with 200 unit cells of length (around 85nm) and different probability distributions for the edge site occupancy. We have calculated the averaged transmission and thermopower, considering an uniform probability distribution, and taking into account around 2000 realizations for each different molecular distribution.  The disorder parameter of the adopted model is defined as the probability occupation of the edge sites of the armchair nanoribbon, simulating a typical experimental setup. 
 
Results for the electronic transmission as a function of the chemical potential are shown in Fig. 4  for an  N= 8 AGNR with random distributions of attached molecules of different lengths (L = 1, 4, and 8) and for several disorder parameters (from 0.5$\%$ up to 20$\%$).  A remarkable fact is that the main Fano antiresonances are hold at the same energy values, centered at the molecule levels, for all molecule distribution considered. However, the transmission at the  eigenenergies of the isolated molecule is not  completely suppressed in the dilute regime ($ < 0.3\%$) due to the fact that the probability of having a sample with no attached molecules is not neglectable. For increasing disorder parameters perfect Fano antiresonances at the molecule spectrum are recovered.
 As expected, in this disordered regime, Anderson localization is manifested and the transmission at the Fermi energy goes to zero.  As it can be observed in the figure, the localization is obtained for a  probability occupation that depends on the molecular length. By increasing the probability occupancy, quantum decoherence effects are more important leading to a overall decreasing in the transmission amplitude except at a particular energy value for which Fano antiresonances are well defined. 
An additional  increase of the disorder parameter leads to a reversal effect on the system localization.

In what follows, we  analyse the effects on thermopower  due to the presence of a random distribution of LPHC molecules attached to the AGNRs. The transmission and  Seebeck coefficients for two different concentrations of Tetracene attached molecules are shown in Fig. 5, for different temperatures.  The upper plots in  (a) $0.5\%$ and (b) $3\%$ display the transmission coefficients and the lower plots show the  Seebeck coefficient as a function of the chemical potential for the corresponding disorder parameter, and different temperatures. The main signatures of the Seebeck coefficients survive  in the presence of the disorder. The thermopower signals are stronger in the less diluted regime [fig. 5(b)] since in this case the Fano antiresonance profiles are better defined. For higher temperatures, as expected, there is a clear enhancement of the thermopower fluctuations, but the main signals are not destroyed indicating the robustness of the system as a sensor device.

\section{Summary}
 We studied the thermopower response of hybrid nanostructured systems composed of graphene nanoribbon conductors with side attached organic-like molecules. The system was described within a Green\' s function formalism and considering a linear regime in the thermoelectric transport calculations. We propose a  molecule sensor based in  thermopower measurements  of  hybrid systems that includes a random distribution of  edge-attached molecules at the nanoribbons considering different occupancy distribution rates. Our previous results\cite{Rosales} which concluded that GNR conductance
  reflects the energy spectrum of these quasi one-dimensional systems are corroborated with the present analysis of the thermopower.  Our present results show that thermopower responses are persistent under temperature and disorder and that may be used as a simple tool to reveal the  nature of the foreign entities attached at the edges of a  graphene  nanoribbon hybrid.

\ack The authors acknowledge Chilean FONDECYT grants 1100560 (PO), 11090212 (LR), and 1100672 (MP), DGIP/USM grants 11.12.17 (LR) and 11.11.62 (MP). AL thanks Brazilian agencies FAPERJ( under grant E-26/101522/2010), CNPq, and the Instituto Nacional de  Ci\^encia e Tecnologia em Nanomateriais de Carbono.

 
\end{document}